\documentclass{article}
\usepackage{amsmath,graphicx}
\usepackage{multirow}
\usepackage{INTERSPEECH2019}
\usepackage[utf8]{inputenc}
\usepackage{url,times,booktabs}
\usepackage{comment}
\usepackage{tabularx}
\usepackage{makecell}
\usepackage{adjustbox}
\usepackage{cite}

\title{End-to-end losses based on speaker basis vectors and \\ all-speaker hard negative mining for speaker verification}
\name{Hee-Soo Heo, Jee-weon Jung, IL-Ho Yang, Sung-Hyun Yoon, Hye-jin Shim, and Ha-Jin Yu$^\dag$\thanks{$^\dag$ Corresponding author}\thanks{This work was supported by the Technology Innovation Program (10076583, Development of free-running speech recognition technologies for embedded robot system) funded By the Ministry of Trade, Industry \& Energy(MOTIE, Korea)}}
\address{School of Computer Science, University of Seoul, South Korea}
\email{zhasgone@naver.com,
jeewon.leo.jung@gmail.com,
heisco@hanmail.net,
ysh901108@naver.com
shimhz6.6@gmail.com,
hjyu@uos.ac.kr}

\begin{document}
\maketitle
\begin{abstract}
In recent years, speaker verification has primarily performed using deep neural networks that are trained to output embeddings from input features such as spectrograms or Mel-filterbank energies. 
Studies that design various loss functions, including metric learning have been widely explored. 
In this study, we propose two end-to-end loss functions for speaker verification using the concept of speaker bases, which are trainable parameters. 
One loss function is designed to further increase the inter-speaker variation, and the other is designed to conduct the identical concept with hard negative mining. 
Each speaker basis is designed to represent the corresponding speaker in the process of training deep neural networks. 
In contrast to the conventional loss functions that can consider only a limited number of speakers included in a mini-batch, the proposed loss functions can consider all the speakers in the training set regardless of the mini-batch composition. 
In particular, the proposed loss functions enable hard negative mining and calculations of between-speaker variations with consideration of all speakers. 
Through experiments on \texttt{VoxCeleb1} and \texttt{VoxCeleb2} datasets, we confirmed that the proposed loss functions could supplement conventional softmax and center loss functions.

\end{abstract}

\section{Introduction}
\label{sec:1}

In recent years, several studies have reported superior results using deep neural networks (DNNs) for extracting speaker embeddings compared to conventional state-of-the-art i-vector-based \cite{dehak2011front} speaker verification systems \cite{d-vector, DeepSpeaker, angularspeaker, generalizedE2E, jung2018complete, jung2018avoiding, snyder2018x}. 
Therefore, several recent studies have mainly focused on designing loss functions to train DNNs to make them suitable for speaker verification. 
Wan \textit{et al.} proposed a generalized end-to-end (GE2E) loss function based on centroids, which are the average embeddings for each speaker, to train DNNs with higher generalization performance \cite{generalizedE2E}. 
Li \textit{et al.} applied a loss function based on angular softmax, which was proposed for face recognition \cite{sphereface}, to create an angular margin between speakers in an embedding space \cite{angularspeaker}.
\\
\indent The conventional studies on loss functions mentioned above do not address the following two problems.
The first problem is that conventional loss functions only consider a limited number of speakers according to mini-batch composition. 
In the process of repeatedly training DNNs with mini-batches of a small size, the parameters of a network could be biased to only the speakers included in one mini-batch. 
The second problem is that excessive overhead occurs in performing hard negative mining, which is important in metric-learning-based loss functions \cite{schroff2015facenet}. 
Hard negative mining is known to have a significant impact on the performance of metric learning. 
However, it is usually performed at regular intervals because of practical issues. 
Ideally, hard negative mining should be conducted for each mini-batch. 
This is because hard negative samples will change as weight parameters are updated every mini-batch. 
Although GE2E has partially solved these problems, there is a limitation that only few speakers can be considered by hard negative mining in GE2E. 

In this paper, we propose loss functions based on speaker bases to handle these problems. 
The speaker bases refer to the column vectors of the weight matrix of the output layer. 
This definition stems from the fact that since the column vectors represent speakers in embedding space (output of the last hidden layer), each column vector can be considered as a basis for that speaker.
This concept can be considered rather a general approach because it can be applied to any DNN-based speaker embedding extraction system that comprises a fully-connected code layer. 
We expect that it would be possible to train all speakers simultaneously and perform hard negative mining in every mini-batch using the loss function based on the speaker bases. 

\section{Related works}
\label{sec:2}

In this section, we introduce various existing loss functions that can be used to train speaker verification systems. 
These loss functions are already successfully applied to speaker verification and the face recognition field.

\subsection{Softmax-based loss function}
The softmax-based loss function is widely used to train DNNs for identification purposes. 
Generally, when the softmax-based loss function is exploited for speaker verification, the output of the last hidden layer is used as the embedding of each utterance after training DNNs. 
The softmax loss function is calculated as:

\begin{equation}
\mathcal{L}_{S} = - \sum\limits_{i=1}^{M}log\frac{exp(W_{y_i}^{T}\boldsymbol{e}_i+b_{y_i})}{\Sigma_{j=1}^{N}exp(W_{j}^{T}\boldsymbol{e}_{i}+b_{j})},
\end{equation}

\noindent where $\boldsymbol{e}_i$ and $y_i$ denote the embedding (output of the last hidden layer) of the $i'th$ utterance and the corresponding speaker label, respectively, M is the number of utterances, $N$ is the number of speakers in the training set, $W=[W_1,W_2,...,W_N]$ and $\boldsymbol{b}=[b_1,b_2,...,b_N]$ are the weight matrix and the bias vector of the output layer, respectively, and $exp(\cdot )$ is the exponential function.

\subsection{Center loss function}
The center loss function was proposed to reduce within-class variations to supplement the softmax-based loss function \cite{centerloss}. 
To reduce within-class variations, loss is calculated based on the mean squared error between the embeddings of each utterance and the center embedding of the corresponding speaker. 
This loss function was successfully applied in the field of face recognition, and high performance improvement was reported. 
The center loss function, defined in equation (2), is used in conjunction with the conventional softmax-based loss function in most cases. 

\begin{equation}
\mathcal{L}_{C} = \frac{\lambda}{2}\sum\limits_{i=1}^{M}||\boldsymbol{e}_i - \boldsymbol{c}_{y_{i}}||_{2}^{2},
\end{equation}

\noindent where $\boldsymbol{c}_i$ is the center embedding of the $i'th$ speaker and $\lambda$ is the weight factor of the center loss function. 
The center embedding of each speaker in the center loss function is not trained based on gradient descent, like other parameters of DNNs. 
Rather, it is trained by moving the center embedding by a scalar $\alpha$ based on the delta center value calculated using the following formula:

\begin{equation}
\Delta\boldsymbol{c}_{k}=\frac{\Sigma_{i=1}^{M}\delta(y_i=k)\cdot(\boldsymbol{c}_{k}-\boldsymbol{e}_i)}{1+\Sigma_{i=1}^{M}\delta(y_i=k)},
\end{equation}

\noindent where $\delta(condition)=1$ if the $condition$ is satisfied; otherwise, $\delta(condition)=0$.

\subsection{Additive margin loss function}
The additive margin softmax (AMsoftmax) loss function was proposed to replace the inner product operation of the softmax-based loss function with the cosine similarity operation \cite{sphereface} and widen the margin between each class in an embedding space \cite{AMsoft}. 
The AMsoftmax loss function is calculated based on the cosine similarity, $cos(\cdot,\cdot )$, so that the embedding between each speaker has an additional margin of $m$, as follows:

\begin{equation}
\mathcal{L}_{AMS} = - \sum\limits_{i=1}^{M}log\frac{exp(s \cdot ( cos(W_{y_i},\boldsymbol{e}_i)-m))}{exp(s \cdot (cos(W_{y_i},\boldsymbol{e}_i)-m))+R_i},
\end{equation}

\begin{equation}
R_i = \Sigma_{j=1,j \neq y_i }^{N}exp(s \cdot cos(W_{y_j},\boldsymbol{e}_i)),
\end{equation}

\noindent where $s$ is a scaling factor for stabilizing the training process of cosine similarity-based loss.
This loss function is an improved version of the angular softmax function \cite{sphereface} and has been successfully applied to speaker recognition systems \cite{angularspeaker, nunes2019additive, yu2019ensemble}.

\subsection{Generalized end-to-end loss function}

The GE2E loss function, a recent one of the advanced versions of triplet loss, was proposed to reduce the distance between the embeddings of each utterance and the centroid embeddings of the corresponding speaker while increasing the distance from the centroid embeddings of other speakers \cite{l2}. 
The most significant characteristic of the GE2E loss function is that it does not calculate the distance between samples but calculates the distance between centroids by averaging the embeddings from the same speaker. 
Wan, Li \textit{et al.} assumed that higher generalization performance could be achieved through a distance comparison with centroid embeddings \cite{l2}. 
For this purpose, the distance between the embedding of the $i'th$ utterance of the $j'th$ speaker and the centroid of the $k'th$ speaker, $S_{ji,k}$, is calculated as follows: 

\begin{equation}
\boldsymbol{S}_{ji,k}=w_{score}\cdot cos(\boldsymbol{e}_{ji},\hat{\boldsymbol{c}}_{k})+b_{score}, 
\end{equation}

\noindent where $w_{score}$ and $b_{score}$ are the trainable parameters for scaling and shifting scores. 
It is important to note that the centroid $\hat{\boldsymbol{c}}_k$, in equation (6), is different from the center embedding, $\boldsymbol{c}_k$, in equation (2). 
Centroid embedding is calculated utilizing the embeddings of each speaker as follows: 

\begin{equation}
\hat{\boldsymbol{c}}_{k}=\frac{1}{M_k}\sum\limits_{i=1}^{M_k}\boldsymbol{e}_{ki},
\end{equation}

\noindent where $M_k$ is the number of utterances of the $k'th$ speaker. 
The GE2E loss function is calculated based on $\boldsymbol{S}_{ji,k}$, as follows: 

\begin{equation}
\mathcal{L}_{G}=\sum\limits_{j,i}1-\sigma(\boldsymbol{S}_{ji,j})+\max_{\substack{ 1<k<N \\ k\neq j}} \sigma(\boldsymbol{S}_{ji,k}),
\end{equation}

\noindent where $\sigma(\cdot)$ is the sigmoid function for stabilizing training. 
In the GE2E loss function, hard negative mining is performed by selecting the largest value among the scores of negative pairs. 
It is important to note that it is required to construct one mini-batch with a few utterances of speakers for calculating the loss defined by equation (6). 
This is because the centroid, $\hat{\boldsymbol{c}}_k$, in equation (6) is calculated from multiple utterances of each speaker. 
This requirement limits the mini-batch configuration, thereby considerably reducing the number of speakers in a mini-batch. 
For example, if one configures a mini-batch of size 100 and each mini-batch includes five utterances per speaker, only 20 speakers are included in a mini-batch. 
This may be too small, considering the number of whole speakers in the dataset for speaker recognition \cite{Voxceleb, martin2010nist}.

\section{Proposed loss functions\\based on class bases}
\label{sec:3}

The core idea behind the proposed loss functions is that the weight matrix between the last hidden layer and output layer in softmax-based loss function can be interpreted as a set of bases where each basis vector represents a class, i.e. a speaker. 
For example, in speaker identification task, to calculate the softmax loss function for 1000 speakers from 128-dimensional embedding, a weight matrix of size [128, 1000] is required. 
This weight matrix can be interpreted as a set of 128-dimensional vectors, which represent each speaker. 
With this interpretation, these basis vectors were trained to replace each speaker's centroid. 
Therefore, it is possible to train the system using the entire classes at every mini-batch regardless of its size. 


We propose an additional loss function to maximize the between-class variation based on the proposed class basis interpretation as the following equation
\begin{equation}
\mathcal{L}_{BC}=\sum\limits_{i=1}^{N}\sum\limits_{j=1,j \neq i}^{N}cos(W_{i},W_{j}),
\end{equation}
\noindent where $N$ is the number of classes, and $W_i$ is the basis of the $i'th$ class.
The conventional loss based on categorical cross entropy also can increase the between class variation. 
However, the conventional loss can consider only the limited number of classes depending on the size of the mini-batch. 
On the other hand, $\mathcal{L}_{BC}$ can consider the entire classes simultaneously. 
We can also interpret the behavior of $\mathcal{L}_{BC}$ as to distribute the bases equally in the embedding space.
We expect that the proposed loss function $\mathcal{L}_{BC}$ is complementary to the conventional center loss function, which only considers within-class variations (see Table 2). 

Hard negative mining, the method to select samples or pairs that are likely to be mistaken in the training set, is known as one of the important factors that determine the performance of metric learning \cite{harwood2017smart}. 
This is because the hard negative samples have a dominant influence on the decision boundaries of the classes. 
In conventional metric learning, hard negative mining is typically performed in a separate phase. 
However, it is difficult to increase the frequency of hard negative mining because of the overhead in the phase. 
The GE2E loss function deals with the same problem in a manner to the proposed loss function, but the number of speakers included in negative mining is quite limited. 
To address this difficulty, we propose another objective function based on the class basis interpretation for performing hard negative mining for all speakers. 
The proposed loss function is defined as:
\begin{equation}
\mathcal{L}_{H}=\sum\limits_{i=1}^{M}\sum\limits_{W_h \in \mathcal{H}_i}log(1+exp(cos(W_h,\boldsymbol{e}_i) - cos(W_{y_i},\boldsymbol{e}_i))),
\end{equation}
\noindent where $\boldsymbol{e}_i$ and $W_{y_{i}}$ denote the $i'th$ utterance and the basis of the corresponding speaker, respectively, and $\mathcal{H}_i$ is the set of the top $H$ speaker bases with large $cos(W_h,\boldsymbol{e}_i)|_{h\neq y_{i}}$ values. 

The main purpose of the loss $\mathcal{L}_{H}$ is to reduce the negative similarities, represented as $cos(W_h,\boldsymbol{e}_i)$, while increasing the positive similarities, represented as $cos(W_{y_i},\boldsymbol{e}_i)$. 
Exponential function $exp(\cdot)$ is applied with the perspective to increase the gradient of the samples with large loss and decrease the gradient of the samples with small loss, simulatneously.
The additional $1+$ term in the $log (x)$ function limits the value of $x$ to greater than one.
This is because the $log (x)$ function has too small a value and an overly large gradient when the value of $x$ is close to zero.
The proposed negative mining objective function enables negative mining on every mini-batch, considering entire classes.

\begin{figure}[t!]
  \centering
  \centerline{\includegraphics[width=0.9\columnwidth]{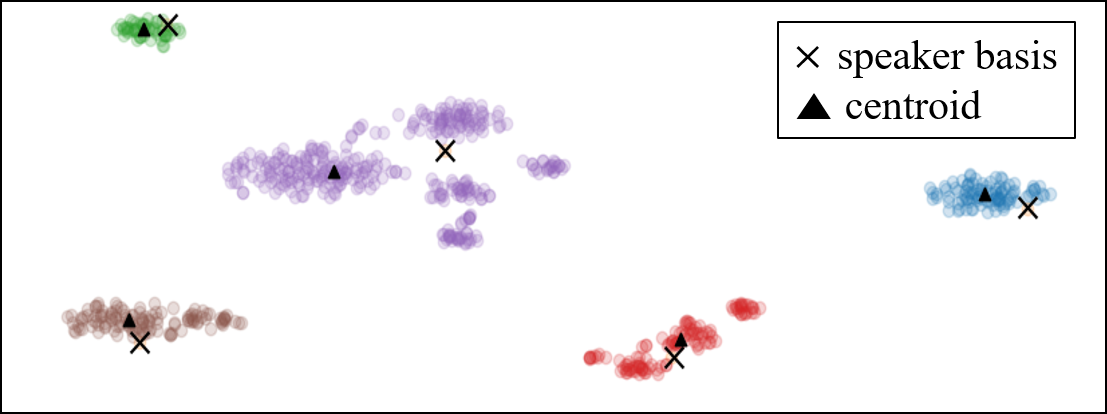}}
  \caption{Embeddings visualized using the t-SNE algorithm \cite{maaten2008visualizing}. Five different colors refer to five randomly selected speakers from the training set. Triangles are the average embeddings of each speaker. $\times$’s are the speaker bases.}
  \label{fig1}
\end{figure}

\begin{figure}[t]
  \centering
  \centerline{\includegraphics[width=0.9\columnwidth]{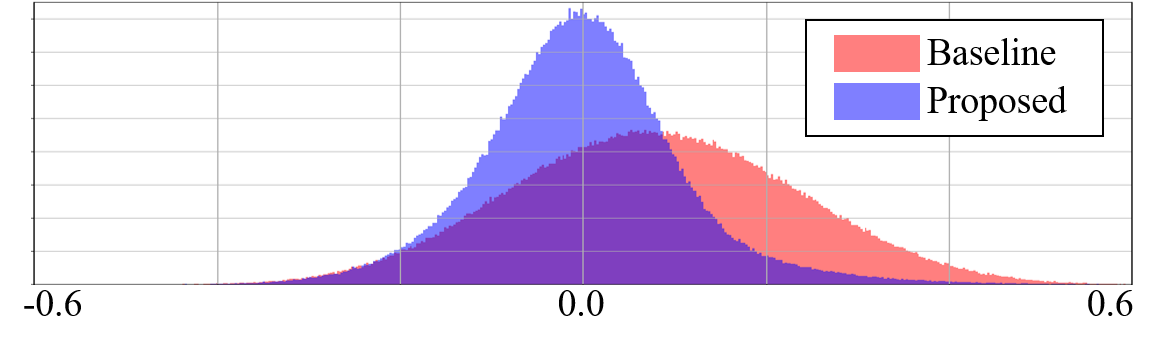}}
  \caption{Histogram of the impostor scores calculated according to the proposed loss function, $\mathcal{L}_{BC}$ ($x$-axis: $cos(\hat{\boldsymbol{c}}_{i}, \hat{\boldsymbol{c}}_{j})|_{i \neq j}$, $y$-axis: $ density$).}
  \label{fig1}
\end{figure}

\begin{table}[t!]
 \renewcommand\thetable{1}
 \caption{DNN architecture ($l$: length of input sequence).}
  \centering
  \label{tab:table1}
  \begin{adjustbox}{width=0.8\columnwidth}
  \begin{tabular}{r c c c}
  
  \Xhline{2\arrayrulewidth}
   layer & output shape  & kernel size  & stride \\
  \hline
  Conv1 & $l \times64\times 16$ & $7\times 7$ &$1\times1$\\
  \hline
  Res1 & $ l \times64\times 16$ & $3\times 3$ &$1\times1$\\
  \hline
  Res2 & $ (l / 2) \times32 \times 32$ & $3\times 3$ &$2\times2$\\
  \hline
  Res3 & $ (l / 4) \times16 \times 64$ & $3\times 3$ &$2\times2$\\
  \hline
  Res4 & $ (l / 8) \times8 \times 128$ & $3\times 3$ &$2\times2$\\
  \hline
  Pool & $ 128$ & Global &Global\\
  
  \Xhline{2\arrayrulewidth}
  \end{tabular}
  \end{adjustbox}
\end{table}

\begin{table*}[h]
 \renewcommand\thetable{2}
 \caption{Performances in terms of EER and hyper-parameters of various systems ($\mathcal{L}_{S}$: softmax-based loss, $\mathcal{L}_{C}$: center loss, $\mathcal{L}_{AMS}$: additive margin softmax, $\mathcal{L}_{G}$: generalized end-to-end loss, $\mathcal{L}_{BC}$: proposed between-speaker variations loss, $\mathcal{L}_H$: proposed hard negative mining loss).}
  \centering
  \label{tab:table1}
  \begin{tabular}{r c c c}
  
  \Xhline{2\arrayrulewidth}
   System & Loss  & hyper-parameters  & EER (\%) \\
  \Xhline{2\arrayrulewidth}
  i-vector PLDA reported in \cite{Voxceleb} & - & - & 8.8 \\
  
  Metric learning reported in \cite{Voxceleb} & - & - & 7.8 \\
  \hline
  Softmax loss (our implementation) & $\mathcal{L}_{S}$ & - & 7.78 \\
  
  Center loss (our implementation)& $\mathcal{L}_{S} + \mathcal{L}_{C}$ & $\lambda=0.001$, $\alpha=0.5$ & \textbf{6.55} \\
  
  AM softmax (our implementation)& $\mathcal{L}_{AMS}$ & $s=5,m=0.35$, weight decay(0.0001) & 7.31 \\
  
  GE2E (our implementation)& $\mathcal{L}_{G}$ & 5 utterances for each speaker, weight decay(0.0001) & 10.65 \\
  \hline
  Proposed 1 & $\mathcal{L}_{S} + \mathcal{L}_{C} + \mathcal{L}_{BC}$ & $\lambda=0.001$, $\alpha=0.5$ & 5.96 \\
  
  Proposed 2 & $\mathcal{L}_{H} + \mathcal{L}_{BC}$ & $H=100$, weight decay(0.0001) & \textbf{5.55} \\
  \Xhline{2\arrayrulewidth}
  \end{tabular}
\end{table*}

Figure 1 illustrates the embeddings, centroids, and speaker bases extracted from the trained DNN using the proposed loss functions ($\mathcal{L}_{BC}+\mathcal{L}_H$), from the utterances of five randomly selected speakers. 
The centroids were calculated by averaging all embeddings of each speaker after the training process. 
Note that these centroids are ideal and shown for comparison purpose only and calculating centroids at every mini-batch is impractical. 
This is because all utterances of a speaker should be considered for calculating one centroid. 
The figure shows that each speaker can be represented by a speaker basis.

Figure 2 shows the histogram of impostor scores, which are calculated as $cos(\hat{\boldsymbol{c}}_{i}, \hat{\boldsymbol{c}}_{j})|_{i \neq j}$, to confirm the effect of the proposed loss function, $\mathcal{L}_{BC}$. 
We compared the difference between the baseline with (trained by $\mathcal{L}_S+\mathcal{L}_C+\mathcal{L}_{BC}$) and without (trained by $\mathcal{L}_S+\mathcal{L}_C$) applying the proposed loss function. 
The result demonstrates that the impostor scores of the training set are reduced by the proposed loss function and between-speaker variations are increased compared to the center loss function (baseline).

\section{Experiments}
\label{sec:4}

We used \texttt{VoxCeleb1} dataset which comprises 1,251 speakers with approximately 330 hours of utterances, following the guideline provided in \cite{Voxceleb}. 
The guideline assigns 1,211 speakers as the training set and 40 speakers as the evaluation set. 
We implemented the DNNs based on Keras with TensorFlow as the back-end \cite{keras, tensorflow,tensorflow2}. 
We used Kaldi for acoustic feature extraction \cite{povey2011kaldi}.

\subsection{Experimental configuration}
\label{ssec:4.3.}

64-dimensional Mel-filterbank energy features were extracted using a 25-ms hamming window with a 10-ms shift \cite{cai2018analysis}. 
Mean normalization was applied over a 3-s sliding window. 
ResNet-34 \cite{he2016deep, he2016identity} was modified as shown in Table 1 and used for extracting 128-dimensional speaker embeddings. 
We used leaky rectified linear unit \cite{leaky} as the activation function. 
The Adam optimizer with a learning rate of 0.001 was utilized with a mini-batch size of 100. 
In our experiments, the performance with the loss function defined by the inner product was degraded by weight decay whereas the performance with the loss function defined by the cosine similarity was improved by weight decay. 
Table 2 demonstrates the hyper-parameters and the EER of baseline and proposed systems. 

\subsection{Results and analysis}
\label{ssec:4.3.}

Results of the implementation in this study showed that the DNN trained using the center loss showed the lowest EER among other losses discussed in Section 2. 
The GE2E loss function, which is expected to have high performance, exhibited a relatively high EER. 
This result is interpreted as a phenomenon caused by a fixed mini-batch size owing to practical issues such as GPU memory. 
In particular, if the mini-batch size was fixed at 100 and five utterances for each speaker were included, one mini-batch would contain only 20 speakers. 
This is an extremely limited number considering the number of all speakers. 
Based on the center loss function, which showed the highest performance among the conventional loss functions, we applied the proposed loss function and compared the performances. 
First, the loss function, $\mathcal{L}_{BC}$, defined using equation (9) with the center loss function could reduce the EER by approximately 9\% relatively. 
This result indirectly shows that between-speaker variations were increased by the proposed loss function, $\mathcal{L}_{BC}$. 
In addition, the error was reduced by 6\% by replacing the center loss function by the proposed loss function, $\mathcal{L}_{H}$. 
Finally, the proposed loss function reduced the error by 15\% relatively.
Based on these results, we found that it is possible to design an effective loss function for speaker verification with the proposed speaker bases.

\begin{table}[b!]
 \renewcommand\thetable{3}
 \caption{DNN architecture in the experiments on \texttt{VoxCeleb2} ($l$: length of input sequence).}
  \centering
  \label{tab:table1}
  \begin{adjustbox}{width=0.8\columnwidth}
  \begin{tabular}{r c c c}
  
  \Xhline{2\arrayrulewidth}
   layer & output shape  & kernel size  & stride \\
  \hline
  Conv1 & $l \times32\times 64$ & $7\times 7$ &$1\times2$\\
  \hline
  Res1 & $ l \times32\times 64$ & $3\times 3$ &$1\times1$\\
  \hline
  Res2 & $ (l / 2) \times16 \times 128$ & $3\times 3$ &$2\times2$\\
  \hline
  Res3 & $ (l / 4) \times8 \times 256$ & $3\times 3$ &$2\times2$\\
  \hline
  Res4 & $ (l / 8) \times4 \times 512$ & $3\times 3$ &$2\times2$\\
  \hline
  Pool & $ 512$ & Global &Global\\
  \hline
  Dense & $ 1024$ & $ 512 \times 1024$ & - \\
  
  \Xhline{2\arrayrulewidth}
  \end{tabular}
  \end{adjustbox}
\end{table}

\subsection{Additional validation using \texttt{VoxCeleb2}}
We conducted experiments using the \texttt{VoxCeleb2} dataset to further verify the proposed loss functions. 
In this configuration, \texttt{VoxCeleb2}, with 6,112 speakers, is used as training set, and the entire \texttt{VoxCeleb1}, with 1,251 speakers, is used as test set. 
The DNNs trained by the proposed loss functions were evaluated using three different trials lists introduced in \cite{Chung18bvoxceleb2}: (1) the original trials list of \texttt{VoxCeleb1}, (2) \texttt{VoxCeleb1-E}, and (3) \texttt{VoxCeleb1-H}. 
As the amount of training increased, we modified the architecture of DNNs as shown in Table 3. 
The b-vector system was implemented for back-end scoring instead of cosine similarity \cite{b-vector}. 
In this system, a b-vector is extracted from two embeddings using element-wise binary operations such as addition, subtraction, and multiplication. 
A 3072-dimensional $(1024 \times 3)$ b-vector is extracted from two 1024-dimensional embeddings. 
The b-vector system was first introduced for scoring the i-vectors. 
Then, it has been shown that the b-vector based system could be efficient classifier of embeddings from DNNs in speaker verification task \cite{jung2018complete, heo2016advanced}. 
In our experiments, the b-vector based system comprises three fully-connected layers, each with 512 nodes, and an output layer with two nodes. 
The two nodes of the output layer indicate whether the two embedding inputs are from the same speaker or not, respectively. 

Table 4 shows the results of an additional validation. 
For the validation, we compared the performances of the system trained by $\mathcal{L}_{S}$ (the best performance among the conventional losses) and the proposed $\mathcal{L}_{BC} + \mathcal{L}_{H}$. 
Experimental results demonstrate that the proposed loss can replace the conventional loss while reducing the error by 21 \% relatively.

\begin{table}[t!]
 \renewcommand\thetable{4}
 \caption{Performance comparison of the experiments using \texttt{VoxCeleb2} ($\mathcal{L}_{S}$: softmax-based loss, $\mathcal{L}_{C}$: center loss, $\mathcal{L}_{BC}$: proposed between-speaker variations loss, $\mathcal{L}_H$: proposed hard negative mining loss, org\texttt{Vox1}: original test lists of \texttt{VoxCeleb1}, \texttt{Vox1-E}: extended test lists based on all speakers of \texttt{VoxCeleb1}, \texttt{Vox1-H}: extended and hard test lists based on all speakers of \texttt{VoxCeleb1}).}
  \centering
  \label{tab:table1}
  \begin{adjustbox}{width=1\columnwidth}
  \begin{tabular}{r c c c}
  
  \Xhline{2\arrayrulewidth}
  Loss & org\texttt{Vox1}  & \texttt{Vox1-E}  & \texttt{Vox1-H} \\
  \Xhline{2\arrayrulewidth}
  $\mathcal{L}_{S} + \mathcal{L}_{C}$ & 2.99 & 3.01 & 5.19\\
  \hline
  $\mathcal{L}_{BC} + \mathcal{L}_{H} (H=100)$ & 2.75 & 2.71 & 4.53\\
  $\mathcal{L}_{BC} + \mathcal{L}_{H} (H=50)$ & \textbf{2.66} & \textbf{2.43} & \textbf{4.08}\\

  \Xhline{2\arrayrulewidth}
  \end{tabular}
  \end{adjustbox}
\end{table}

\section{Conclusions}
\label{sec:5}

In this study, we proposed an interpretation of the weight matrix of the output layer as class bases. 
We applied the proposed end-to-end loss functions using this interpretation for speaker verification. 
The proposed loss function comprises $\mathcal{L}_{BC}$ for increasing between-speaker variations and $\mathcal{L}_{H}$ for hard negative mining. 
The biggest advantage of the proposed loss functions is that regardless of the composition of the mini-batch, all speakers can be considered simultaneously. 
The experimental results obtained for the \texttt{VoxCeleb} showed that the error in the proposed loss function was reduced by approximately 15\% relatively compared with the error in the conventional loss functions. 
In addition, we found that the proposed loss function could replace the conventional loss functions. 

In this paper, we intensively analyzed the performances depending on the various loss functions without modification of the DNN architecture. 
As a future work, we plan to apply the proposed loss functions to various DNN architectures including the x-vector system \cite{snyder2018x}. 
In addition, we also have a plan to apply the proposed interpretation regarding the class basis of the output layer into other domain tasks such as face recognition. 

\vfill\pagebreak

\newpage\newpage
\bibliographystyle{IEEEtran}
\bibliography{mybib}

\end{document}